\documentclass[preprint,aps,floats]{revtex4}
\input epsf.tex
\def\DESepsf(#1 width #2){\epsfxsize=#2 \epsfbox{#1}}
\def\be {\begin{equation}}
\def\ee {\end{equation}}

\def\ra{\rightarrow}

\def\l {\lambda}

\def\betak {B\rightarrow \eta K}
\def\betak0 {B^{0}\rightarrow \eta' K^{0}}

\def\betapkstr0 {B^{0}\rightarrow \eta' K^{*0}}
\def\bpetak0 {B^{0}\rightarrow \eta' K^{0}}

\def\msnus {m_{\tilde\nu_{iL}}^2}
\def\msells {m_{\tilde e_{iL}}^2}

\def\lappeq{\mathrel{\rlap{\raise.5ex\hbox{$<$}}
                    {\lower.5ex\hbox{$\sim$}}}}
       
\def\beq{\begin{equation}} \def\eeq{\end{equation}}
\def\beqa{\begin{eqnarray}} \def\eeqa{\end{eqnarray}}
\def\ba{\begin{array}} \def\ea{\end{array}}

\begin{document}
%\preprint{\vbox{\hbox{}\hbox{KEK-TH-9xx}}}
\draft
\title{CP Asymmetries of $B \to \phi K$ and $B \to \eta^{(\prime)} K$ Decays\\
Using a Global Fit in QCD Factorization}

\author{ Bhaskar Dutta$^{1}$\footnote{duttabh@uregina.ca},
~~ C. S. Kim$^{2}$\footnote{cskim@yonsei.ac.kr},
~~ Sechul Oh$^{3}$\footnote{scoh@post.kek.jp}, ~~ and ~~
Guohuai Zhu$^{3}$\footnote{zhugh@post.kek.jp}  }

\affiliation{
$^1$Department of Physics, University of Regina, SK, S4S 0A2, Canada \\
$^2$Department of Physics, Yonsei University, Seoul
120-479, Korea \\
$^3$Theory Group, KEK, Tsukuba, Ibaraki 305-0801, Japan}

\begin{abstract}
\noindent We analyze the CP asymmetries of $B\to \phi K$ and $B\to
\eta^{(\prime)} K$ modes in the QCD improved factorization
framework. For our calculation we use the phenomenological
parameters predetermined from the global fit for the available
$B\to PP$ and $VP$ modes (without the quark-level subprocess $b\to
s\bar s s$).  We show that the large negative $\sin
(2\phi_1)_{\phi K_s}$ and the large branching ratio for
$B^{\pm}\to \eta' K^{\pm}$ can be simultaneously explained in the
context of supersymmetry (SUSY). The R-parity conserving SUGRA
models are used and their parameter space is constrained with the
observed dark matter relic density along with other experimental
constraints. The R-parity violating SUSY models are also used to
show that they can provide solutions. We calculate the CP
asymmetries for different $B^{\pm (0)} \to \phi K^{\pm (0)}$ and
$B^{\pm (0)} \to \eta^{(')} K^{\pm (0)}$ modes and show that the
SUSY model predictions are consistent with the available
experimental data.
\end{abstract}
\maketitle

%%%%%%%%%%%%%%%%%%%%%%%%%%%%%%%%%%%%%%%%%%%%%%%%%%%%%%%%%%%%%%%%%%%%%
%%%%%%%%%%%%%%%%%%%%%%%%%%%%%%%%%%%%%%%%%%%%%%%%%%%%%%%%%%%%%%%%%%%%%

Recent results from Belle and BaBar on $B$ decays  provide an opportunity
to test the source of CP violation. The standard model (SM) source for CP
violation arises from the CKM matrix which has only  one phase. The SM
predictions for CP asymmetry for different $B$ decay modes are being tested
at the $B$ factories and discrepancy seems to be emerging.
For example, the modes $B\to \phi K_s$ and $B\to J/\Psi K_s$ are uniquely
clean in their theoretical interpretations.  Among these decay modes,
$B\to \phi K_s$ occurs only at one loop level in the SM and hence is a
very promising mode to see the effects of new physics.  In the SM, it
is predicted that the CP asymmetries of $B\to \phi K_s$ and $B\to J/\Psi
K_s$ should measure the same $\sin(2 \phi_1)$ with negligible
${\cal O}(\lambda^2)$ difference \cite{worah}.
The Belle and BaBar experiments measure: \cite{belbab02}
\begin{equation}
\label{eq01}
\sin(2 \phi_1)_{J/\Psi K_s}= 0.734 \pm 0.055,
\end{equation}
and \cite{tom, belle03}
\begin{equation}
\label{eq02}
\sin(2 \phi_1)_{\phi K_s} =  -0.96\pm 0.5^{+0.09}_{-0.11} ~({\rm Belle}),
 ~~~ 0.45\pm 0.43\pm 0.07 ~({\rm BaBar}).
\end{equation}
The world average shows a $2.7\sigma$ disagreement between
$\sin(2 \phi_1)_{\phi K_s}$ and $\sin(2 \phi_1)_{J/\Psi K_s}$.
The $\sin(2 \phi_1)_{J/\Psi K_s}$ being a tree level process is in
excellent agreement with the SM theoretical prediction,
$\sin(2 \phi_1)_{\rm SM}=0.715^{+0.055}_{-0.045}$ \cite{Buras}.

Experimental data is also available for the decay modes $B^{\pm( 0)}
\to \eta^{(\prime)} K^{\pm (0)}$ which involve the quark-level subprocess
$b \to s \bar s s$ as in $B\to \phi K_s$.
The world average value of the measured branching ratio (BR) is
${\cal B}(B^{\pm}\to \eta' K^{\pm})= (77.6 \pm 4.6) \times 10^{-6}$
\cite{Gordon:2002yt,Aubert:2003bq,Cronin-Hennessy:kg},
which is larger than the predicted SM value.
The results also exist for $\sin(2 \phi_1)_{\eta' K_s}= 0.33 \pm 0.34$
\cite{HamelDeMonchenault:2003pu,Abe:2002np} and the CP rate asymmetries
$\mathcal{A}_{\rm CP}$ for different $B^{\pm} \to \phi K^{\pm}$ and
$B^{\pm} \to \eta^{(')} K^{\pm}$ modes \cite{Aubert:2003tk}.

In this letter we try to find a \emph{consistent} explanation for
\emph{all the observed data} in charmless hadronic $B \to PP$ and
$B \to VP$ decays [$P(V)$ denotes a pseudoscalar (vector) meson]
in the framework of QCD factorization (QCDF). We calculate the BRs
and the CP asymmetries for the decay processes $B^{\pm (0)} \to
\phi K^{\pm (0)}$ and $B^{\pm (0)} \to \eta^{(')} K^{\pm (0)}$ in
the SM and its SUSY extensions with R-parity conservation (SUGRA
models), and with R-parity violation. The required QCDF input
parameters for the calculation are determined by using a global
fit of all possible $B$ decay modes without the subprocess $b \to
s \bar s s$, since this process may involve new physics.

Previously, new interactions were invoked to explain the large BRs
of $B^{\pm (0)} \to \eta' K^{\pm (0)}$ and the large negative
$\sin(2\phi_1)_{\phi K_s}$
\cite{new1,we,new2,khalil,Arnowitt:2003ev,dkoz}. However, attempts
of simultaneous explanation for the observed data of these decay
modes were made only by using the naive factorization technique
\cite{we,khalil}. We adopt the newly developed QCD improved
factorization \cite{bbns} for the calculation in this work.  This
approach allows us to include the possible non-factorizable
contributions. In the heavy quark limit $m_b >> \Lambda_{\rm
QCD}$, the hadronic matrix element for $B \to M_1 M_2$ due to a
particular operator $O_i$ can be written in the QCDF as follows:
\begin{eqnarray}
\langle M_1 M_2 | O_i | B \rangle
 = \langle M_1 M_2 | O_i | B \rangle_{\rm NF}
 \cdot \left[ 1 + \sum_n r_n (\alpha_s)^n
 + O \left( {\Lambda_{\rm QCD} \over m_b} \right) \right],
\end{eqnarray}
where NF denotes the naive factorization.  The second and third term
in the square bracket represent the radiative corrections in $\alpha_s$
and the power corrections in $\Lambda_{\rm QCD} / m_b$.
The decay amplitudes for $B \to M_1 M_2$ can be expressed as
\begin{eqnarray}
 {\cal A}(B \to M_1 M_2) = {\cal A}^f(B \to M_1 M_2)
 + {\cal A}^a(B \to M_1 M_2)~,
\end{eqnarray}
where
\begin{eqnarray}
 {\cal A}^f (B \to M_1 M_2) &=& \frac{G_{F}}{\sqrt{2}}
    \sum\limits_{p=u,c} \sum\limits_{i=1}^{10} V_{pb} V^*_{pq}
    ~ a_{i}^{p} \langle M_1 M_2 {\vert} O_{i} {\vert} B
    \rangle_{\rm NF}~, \nonumber \\
 {\cal A}^a (B \to M_1 M_2) &=& \frac{G_{F}}{\sqrt{2}}~
    f_{B} f_{M_1} f_{M_2} \sum\limits_{p=u,c} \sum\limits_{i=1}^{10}
    V_{pb} V^*_{pq}~ b_{i} ~.
\end{eqnarray}
Here ${\cal A}^f (B \to M_1 M_2)$ includes vertex corrections, penguin
corrections, and hard spectator scattering contributions which are
absorbed into the QCD coefficients $a_i$, and ${\cal A}^a (B \to M_1 M_2)$
includes weak annihilation contributions which are absorbed into the
parameter $b_i$. The explicit expressions of $a_i$ and $b_i$ can be found
in Refs. \cite{dkoz,bbns,Du:2001hr}.
The relevant end-point divergent integrals are parametrized as \cite{bbns}
\begin{eqnarray}
X_{H,A} \equiv \int^1_0 {dx \over x}
 \equiv \left( 1 + \rho_{H,A} e^{i \phi_{H,A}} \right)
 {\rm ln} {m_B \over \Lambda_h} ~,
\label{XAXH}
\end{eqnarray}
where $X_H$ and $X_A$ denote the hard spectator scattering contribution and
the annihilation contribution, respectively.
Here the phases $\phi_{H,A}$ are arbitrary, $0^0 \leq \phi_{H,A} \leq 360^0$,
and the parameter $\rho_{H,A} \leq 1$, and the scale $\Lambda_h = 0.5$ GeV
assumed phenomenologically \cite{bbns}.
In principle, the parameters $\rho_{H,A}$ and $\phi_{H,A}$ for $B \to PP$
decays can be different from those for $B \to VP$ decays.

We use the global analysis to determine the QCD parameters $X_H$ and $X_A$
\cite{dkoz}. In the global analysis we exclude the decay modes
whose (dominant) internal quark-level process is $b \to s \bar s s$:
for example, $B \to \phi K$ and $B \to \eta^{(\prime)} M$, where $M$
denotes a light meson, such as $\pi, ~ K, ~ \rho, ~ K^*$.
The reason for such a global analysis is due to the fact that the
$b \to s \bar s s$ mode may require the existence of new physics.
We  use twelve $B \to PP$ and $VP$ decay modes, including $B \to \pi \pi$,
$\pi K$, $\rho \pi$, $\rho K$, $\omega \pi$, $\omega K$, to determine the
QCD parameters, $\rho_{A,H}^{PP,PV}$ and $\phi_{A,H}^{PP,PV}$.
The parameters are distinguished by their superscripts $P$ and $V$ which
denote final state mesons.
Now, if $\rho_{A,H}$ and $\phi_{A,H}$ are large, the effects of $X_{A,H}$
can be large and give rise to a large annihilation (and/or hard spectator
scattering) contribution, which indicates
the theory becomes less reliable and suspicious due to that large non-perturbative
contribution.
We find that it is possible to obtain a good global fit ($\chi^2_{min}$=7.5)
with large $X_{A,H}$ effects.
In this scenario, the BR for $B^+ \to \eta' K^+$ is $74.7 \times 10^{-6}$,
which saturates the experimental limit with the very large $X_{A,H}$
effects \cite{dkoz} and the BR for $B^+ \to \phi K^+$ is small:
$4.0 \times 10^{-6}$.
It is possible to obtain successful fits to all $B\to \phi K$ and
$B\to \eta' K$ data in our SUGRA models.  But, since the effects of
$X_{A,H}$ are large, we will just comment on the fits in this scenario in
our result sections.

In Ref. \cite{dkoz}, we generated a fit (with $\chi^2_{min} =18.3$) with
the (relatively) small $X_A$ and $X_H$ effects.
The corresponding theoretical inputs for this fit are as follows (we will
use these values in this work):
\begin{eqnarray}
&\mbox{}& \lambda =0.2198, ~~  A=0.868, ~~ \phi_3 =86.8^0, ~~
  |V_{ub}|=3.35 \times 10^{-3},
\nonumber \\
&\mbox{}& \mu =2.1 ~{\rm GeV}, ~~ m_s (2 ~{\rm GeV}) =85 ~{\rm MeV},
 ~~ f_B =220 ~{\rm MeV},
\nonumber \\
&\mbox{}& F^{B \pi} =0.249, ~~  R_{\pi K} =1, ~~ A^{B \rho} =0.31,
\nonumber \\
&\mbox{}&  \rho^{PP}_A =0, ~~ \rho^{VP}_A =0.5,
 ~~ \rho^{PP}_H =1, ~~ \rho^{VP}_H =0.746,
\nonumber \\
&\mbox{}& \phi^{VP}_A =-6^0, ~~ \phi^{VP}_H = \phi^{PP}_H =180^0.
\label{2ndfitinputs}
\end{eqnarray}
Note that in this case the effect of the weak annihilation parameter
$X_A$ is relatively small ($i.e.$, $\rho^{PP}_A =0$ and $\rho^{VP}_A =0.5$),
and the effect of the hard spectator scattering parameter $X_H$ is also very
small, because $\rho^{PP}_H =1$, $\rho^{VP}_H =0.746$, and
$\phi^{PP}_H = \phi^{VP}_H =180^0$ so that the terms 1 and
$\rho_H e^{i \phi_H}$ in $X_H$ cancel each other in Eq. (\ref{XAXH}).
Based on the above inputs, the BRs and CP asymmetries for $B \to \phi K$
and $B \to \eta' K$ are predicted and shown in Table I.  We see that the
predicted $\mathcal{B}(B^{+(0)} \to \eta' K^{+(0)})$ are smaller than the
measured values and the predicted $\sin(2\phi_1)_{\phi K_s}$ is also very
different from the world average.
The predicted $\mathcal{B}(B^+ \to \eta K^+)$ is smaller than the
experimental data as well.

%%%%%%%%%%%%%%%%%%%%%%%%%%%%%%%%%%%%%%%%%%%%%%%%%%%%%%%%%%%%%%%%%
%%%%%%%%%%%%%%%%%%%%%%%%%%%%%%%%%%%%%%%%%%%%%%%%%%%%%%%%%%%%%%%%%
\begin{table}
\caption{The branching ratios (in unit of $10^{-6}$) and CP asymmetries
of $B \to \phi K$ and $B \to \eta^{(\prime)} K$ decays are shown.
The values in the parenthesis are experimental numbers
\cite{tom, belle03,Gordon:2002yt,Aubert:2003bq,Cronin-Hennessy:kg,
HamelDeMonchenault:2003pu,Abe:2002np,Aubert:2003tk,Beneke:2003zv}.
Here the inputs for the fit with the small effects of $X_A$ and $X_H$
are used.  The first row of the last column shows the $\sin(2\phi_1)$
for $B \to \phi K_s$ and the last row in the same column shows the
value for $B \to \eta' K_s$.}
\smallskip
\begin{tabular}{c|c|c||c|c|c||c}
\hline \hline
Decay mode & BR & $\mathcal{A}_{\rm CP}$ & Decay mode & BR
& $\mathcal{A}_{\rm CP}$ & $\sin{(2\phi_1)}$
\\ \hline
$B^+ \to \phi K^+$ & 7.3 & 0 & $B^0 \to \phi K^0$ & 6.7
& 0.01 & $\phi K_s$: 0.68
\\
 & ($9.2 \pm 1.0$) & ($0.03 \pm 0.07$) &  & ($7.7 \pm 1.1$)
 & ($0.19 \pm 0.68$) & (Eq. \ref{eq02})
\\
$B^+ \to \eta' K^+ $ & 51 & 0.01 & $B^0 \to \eta' K^0$ & 46.8
& 0.016 & $\eta' K_s$: 0.57
\\
 & ($77.6 \pm 4.6$) & ($0.02 \pm 0.04$) &  & ($60.6 \pm 7$)
 & ($0.8 \pm 0.18$) & ($0.33 \pm 0.34$)
\\
$B^+ \to \eta K^+$ & 1.9 & $-0.16$ & $B^0 \to \eta K^0$ & 1.7
& $-0.16$ &
\\
 & ($3.1 \pm 0.7$) & ($-0.32 \pm 0.20$) &  & ($<4.6$) & ($-$) &
\\ \hline \hline
\end{tabular}
\end{table}
%%%%%%%%%%%%%%%%%%%%%%%%%%%%%%%%%%%%%%%%%%%%%%%%%%%%%%%%%%%%%%%%%
%%%%%%%%%%%%%%%%%%%%%%%%%%%%%%%%%%%%%%%%%%%%%%%%%%%%%%%%%%%%%%%%%

In the following two sections, we will discuss the CP asymmetries and
BRs of $B \to \phi K$ and $B \to \eta' K$ modes in the context of
SUSY models.
\\

%%%%%%%%%%%%%%%%%%%%%%%%%%%%%%%%%%%%%%%%%%%%%%%%%%%%%%
{\bf [1] R-parity violating SUSY case}
%%%%%%%%%%%%%%%%%%%%%%%%%%%%%%%%%%%%%%%%%%%%%%%%%%%%%%

The R-parity violating (RPV) part of the superpotential of
the minimal supersymmetric standard
model can have the following terms
\begin{eqnarray}
 {\cal W}_{\rm RPV} &=& \kappa_iL_iH_2 + \l_{ijk}L_iL_jE_k^c
   + \l'_{ijk}L_iQ_jD_k^c + \l''_{ijk}U_i^cD_j^cD_k^c ~,
\label{superpot}
\end{eqnarray}
where $E_i$, $U_i$ and $D_i$ are respectively the $i$-th type
of lepton, up-quark and down-quark singlet superfields, $L_i$ and $Q_i$ are
the SU$(2)_L$ doublet lepton and quark superfields, and $H_2$ is the Higgs
doublet with the appropriate hypercharge.

For our purpose, we will assume only  $\l'-$type  couplings to be present.
Then, the effective Hamiltonian for charmless hadronic $B$ decay can be
written as \cite{Choudhury:1998wc},
\begin{eqnarray}
{H_{eff}^{\lambda'}} (b\ra \bar d_j d_k d_n)
   &=& d^R_{jkn} \left[ \bar d_{n\alpha} \gamma^\mu_L d_{j\beta}
          ~~ \bar d_{k\beta} \gamma_{\mu R} b_{\alpha} \right]
       + d^L_{jkn} \left[ \bar d_{n\alpha} \gamma^\mu_L b_{\beta}
          ~~ \bar d_{k\beta} \gamma_{\mu R} d_{j\alpha} \right],
\nonumber \\
{H_{eff}^{\lambda'}}(b\ra \bar u_j u_k d_n)
   &=& u^R_{jkn} \left[ \bar u_{k\alpha} \gamma^\mu_L u_{j\beta}
          ~~ \bar d_{n\beta} \gamma_{\mu R} b_{\alpha} \right].
\end{eqnarray}
Here the coefficients $d^{L,R}_{jkn}$ and $u^R_{jkn}$ are defined as
\begin{eqnarray}
d^R_{jkn} &=& \sum_{i=1}^3 {\l'_{ijk}\l'^{\ast}_{in3} \over 8\msnus}, ~~~
d^L_{jkn} =  \sum_{i=1}^3 {\l'_{i3k}\l'^{\ast}_{inj} \over 8\msnus}, ~~
(j,k,n=1,2)  \nonumber \\
u^R_{jkn} &=& \sum_{i=1}^3 {\l'_{ijn}\l'^{\ast}_{ik3} \over 8\msells}, ~~~
(j,k=1, \ n=2)
\end{eqnarray}
where $\alpha$ and $\beta$ are color indices and
$\gamma^\mu_{R, L} \equiv \gamma^\mu (1 \pm \gamma_5)$.
The leading order QCD correction to this operator is given by a scaling
factor $f\simeq 2$ for $m_{\tilde\nu}=200$ GeV.
We refer to Refs. \cite{Choudhury:1998wc,dko2} for the relevant notations.

The RPV SUSY part of the decay amplitude of $B^- \to \phi K^-$ is given by
\begin{eqnarray}
{\cal A}^{RPV}_{\phi K} &=& \left( d^L_{222} + d^R_{222} \right)
 \tilde a A_{\phi}~,
\end{eqnarray}
where the coefficient $\tilde a$ is expressed as
\begin{eqnarray}
\tilde a &=& {1 \over N_c} \left[ 1 - {C_F \alpha_s \over 4 \pi}
 \left( V_{\phi} +12 - \left[ {4  \over 3} \ln{m_b \over \mu}
 - G_{\phi}(0) \right] +{4 \pi^2 \over N_c} H(BK, \phi) \right) \right].
\end{eqnarray}
It has been noticed \cite{dko2} that the RPV part of
the decay amplitude for $B \to \eta' K$,
${\cal A}^{\rm RPV}_{\eta' K}$, is
proportional to $( d^L_{222} -d^R_{222})$, while the RPV part of
the decay amplitude for $B \to \phi K$,
${\cal A}^{\rm RPV}_{\phi K}$,
is proportional to $( d^L_{222} +d^R_{222})$.
It has been also pointed out \cite{dko2} that the opposite relative
sign between $d^L_{222}$ and $d^R_{222}$ in the modes $B \to \eta' K$
and $B \to \phi K$ appears due to the different parity in the final
state mesons $\eta'$ and $\phi$, and this different combination of
$( d^L_{222} -d^R_{222})$ and $( d^L_{222} +d^R_{222})$ in these
modes plays an important role to explain both the large BRs for
$B \to \eta' K$ and the large negative value of
$\sin(2 \phi_1)_{\phi K_s}$ at the same time.

We define the new coupling terms $d^L_{222}$ and $d^R_{222}$ as follows:
\begin{eqnarray}
d^L_{222} \propto | \lambda^{\prime}_{i32}
\lambda^{\prime *}_{i22}| e^{i \theta_L}~, ~~
d^R_{222} \propto |\lambda^{\prime}_{i22} \lambda^{\prime *}_{i23}|
e^{i \theta_R},
\end{eqnarray}
where $\theta_L$ and $\theta_R$ denote new weak phases of the product
of new couplings
$\lambda^{\prime}_{i32} \lambda^{\prime *}_{i22}$ and
$\lambda^{\prime}_{i22} \lambda^{\prime *}_{i23}$, respectively, as defined
by $\lambda^{\prime}_{332} \lambda^{\prime *}_{322} \equiv
| \lambda^{\prime}_{332} \lambda^{\prime *}_{322}| e^{i \theta_L}$ and
$\lambda^{\prime}_{322} \lambda^{\prime *}_{323} \equiv
|\lambda^{\prime}_{322} \lambda^{\prime *}_{323}| e^{i \theta_R}$.

We consider two different cases as follows.

{\bf Case (a):} $d^L_{222} \neq 0$ and $d^R_{222}=0$, $i.e.$,
\begin{eqnarray}
 |\lambda^{\prime}_{322}| &=& 0.077 ~,~ |\lambda^{\prime}_{332}| = 0.077 ~,~
 |\lambda^{\prime}_{323}| =0~, \nonumber \\
 \theta_L &=& 1.5 ~,~  m_{\rm SUSY} = 200 ~{\rm GeV}.
\label{rpv1}
\end{eqnarray}
Our results are summarized in Table II.
We use the negative $d^L_{222}$ and the scaling factor \cite{dko2}.
We find that $\sin(2 \phi_1)_{\phi K_s}$ can be brought down to 0 at most.
The BRs, $\mathcal{B}(B^+ \to \eta K^+)$ and $\mathcal{B}(B^{0} \to
\eta K^0)$, are larger compared to the experimental values, but the
experiments (Belle, BaBar and CLEO) are not quite in agreement and the
experimental errors are also large for these modes.
In Table II, we used  $\delta' =0$, but if we use $\delta' =30^0$, the
$\sin(2\phi_1)_{\phi K_s}$ can be larger negative: $-0.2$, after satisfying
all the constraints [especially the $\mathcal{B}(B^+ \to \eta K^+)$ and the
$\sin(2 \phi_1)_{\eta' K_s}$].

%%%%%%%%%%%%%%%%%%%%%%%%%%%%%%%%%%%%%%%%%%%%%%%%%%%%%%%%%%%%%%%%%
%%%%%%%%%%%%%%%%%%%%%%%%%%%%%%%%%%%%%%%%%%%%%%%%%%%%%%%%%%%%%%%%%
\begin{table}
\caption{{\bf Case (a):} the branching ratios (in unit of $10^{-6}$)
and CP asymmetries of $B \to \phi K$ and $B \to \eta^{(\prime)} K$ decays
are calculated in the framework of R-parity violating SUSY.}
\smallskip
\begin{tabular}{c|c|c||c|c|c||c}
\hline \hline
Decay mode & BR & $\mathcal{A}_{\rm CP}$ & Decay mode & BR
& $\mathcal{A}_{\rm CP}$ & $\sin(2\phi_1)$
\\ \hline
 $B^+ \to \phi K^+$ & 8.9 & $-0.17$
 & $B^0 \to \phi K^0$ & 8.2 & $-0.18$ & $\phi K_s$: $-0.03$
\\
 $B^+ \to \eta' K^+$ & 72.0 & 0.16
 & $B^0 \to \eta' K^0 $ & 66.0 & 0.16 & $\eta' K_s$: $-0.2$
\\
 $B^+ \to \eta K^+$ & 10.5 & 0.25
 & $B^0 \to \eta K^0 $ & 9.7 & 0.25 &
\\ \hline \hline
\end{tabular}
\end{table}
%%%%%%%%%%%%%%%%%%%%%%%%%%%%%%%%%%%%%%%%%%%%%%%%%%%%%%%%%%%%%%%%%
%%%%%%%%%%%%%%%%%%%%%%%%%%%%%%%%%%%%%%%%%%%%%%%%%%%%%%%%%%%%%%%%%

{\bf Case (b):} $d^L_{222} \neq d^R_{222} \neq0$, i.e.,
\begin{eqnarray}
 |\lambda^{\prime}_{322}| &=& 0.076~,~ |\lambda^{\prime}_{332}| = 0.076~,~
 |\lambda^{\prime}_{323}| = 0.064~, \nonumber \\
 \theta_L &=& 1.32 ~,~ \theta_R = -1.29 ~,~~ m_{\rm SUSY} = 200 ~{\rm GeV}.
\label{rpv}
\end{eqnarray}
Our results are summarized in Table III.
We find that $\sin(2 \phi_1)_{\phi K_s}$ can be large negative.
In addition to the parameters given in Eq. (\ref{rpv}), we also used the
additional strong phase $\delta' =30^0$, which can arise from the power
contributions of $\Lambda_{QCD}/ m_b$ neglected in the QCDF scheme,
and whose size can be in principle comparable to the strong phase arising
from the radiative corrections of $O(\alpha_s)$.
If $\delta' =0$ is used, we obtain $\sin(2\phi_1)_{\phi K_s}= -0.2$.

%%%%%%%%%%%%%%%%%%%%%%%%%%%%%%%%%%%%%%%%%%%%%%%%%%%%%%%%%%%%%%%%%
%%%%%%%%%%%%%%%%%%%%%%%%%%%%%%%%%%%%%%%%%%%%%%%%%%%%%%%%%%%%%%%%%
\begin{table}
\caption{{\bf Case (b):} the branching ratios (in unit of $10^{-6}$)
and CP asymmetries of $B \to \phi K$ and $B \to \eta^{(\prime)} K$ decays
are calculated in the framework of R-parity violating SUSY.}
\smallskip
\begin{tabular}{c|c|c||c|c|c||c}
\hline \hline
Decay mode & BR & $\mathcal{A}_{\rm CP}$ & Decay mode & BR
& $\mathcal{A}_{\rm CP}$ & $\sin(2\phi_1)$
\\ \hline
 $B^+ \to \phi K^+$ & 10.2 & $-0.05$
 & $B^0 \to \phi K^0$ & 9.5 & $-0.04$ & $\phi K_s$: $-0.61$
\\
 $B^+ \to \eta' K^+$ & 74.0 & 0.11
 & $B^0 \to \eta' K^0 $ & 67.7 & 0.11 & $\eta' K_s$: 0.48
\\
 $B^+ \to \eta K^+$ & 6.7 & 0.06
 & $B^0 \to \eta K^0 $ & 6.1 & 0.06 &
\\ \hline \hline
\end{tabular}
\end{table}
%%%%%%%%%%%%%%%%%%%%%%%%%%%%%%%%%%%%%%%%%%%%%%%%%%%%%%%%%%%%%%%%%
%%%%%%%%%%%%%%%%%%%%%%%%%%%%%%%%%%%%%%%%%%%%%%%%%%%%%%%%%%%%%%%%%

In the case of the large $X_{A,H}$ effects with $\chi^2_{\min}=7.5$
where the BR of $B^+ \to \eta' K^+$ is large, we can use the R-parity
violating SUSY couplings to raise the BR of $B^+ \to \phi K^+$ (which
is small, $4.0 \times 10^{-6}$ to begin with).
It is possible to raise ${\cal B}(B^+ \to \phi K^+)$ to $(8-9) \times
10^{-6}$, but its CP rate asymmetry $\mathcal{A}_{\rm CP}(B^+ \to
\phi K^+)$ is large $\sim -0.4$ and $\sin{(2\phi_1)}_{\phi K_s}$ can be
brought down to at most $-0.16$.

The RPV terms can arise in the context of SO(10) models which explain the
small neutrino mass and has an intermediate breaking scale where $B-L$
symmetry gets broken by $(16+\bar{16})$ Higgs. These additional Higgs form
operators like $16_H16_m16_m16_m/M_{pl}$ ($16_m$ contains matter fields)
and generate the RPV terms \cite{Mohapatra:1996pu}.

%%%%%%%%%%%%%%%%%%%%%%%%%%%%%%%%%%%%%%%%%%%%%%%%%%%%%%
{\bf [2] R-parity conserving SUSY case}
%%%%%%%%%%%%%%%%%%%%%%%%%%%%%%%%%%%%%%%%%%%%%%%%%%%%%%

As an example of the R-parity conserving (RPC) SUSY case,
we will consider the supergravity
(SUGRA) model with the simplest possible non-universal soft terms which
is the simplest extension of the minimal SUGRA (mSUGRA) model. In this
model the lightest SUSY particle is stable and this particle can explain
the dark matter content of the universe.  The recent WMAP result provides
\cite{wmap}:
\begin{equation}
\Omega_{\rm CDM} h^2 =0.1126^{+0.008}_{-0.009},
\end{equation}
and we implement 2$\sigma$ bound in our calculation.

In the SUGRA model, the superpotential and soft SUSY breaking terms
at the grand unified theory (GUT) scale are given by
\begin{eqnarray}
{\cal W} &=& Y^U Q H_2 U + Y^D Q H_1 D + Y^L L H_1 E + \mu H_1 H_2 ,
\nonumber \\
{\cal L}_{\rm soft} &=& - \sum_i m_i^2 |\phi_i|^2
 - \left[ {1 \over 2} \sum_{\alpha} m_{\alpha} \bar \lambda_{\alpha}
 \lambda_{\alpha} + B \mu H_1 H_2 \right.  \nonumber \\
&\mbox{}& \left. + (A^U Q H_2 U + A^D Q H_1 D + A^L L H_1 E)
  + {\rm H.c.} \right],
\end{eqnarray}
where $E$, $U$ and $D$ are respectively the lepton, up-quark and
down-quark singlet superfields, $L$ and $Q$ are the SU$(2)_L$ doublet
lepton and quark superfields, and $H_{1,2}$ are the Higgs doublets.
$\phi_i$ and $\lambda_{\alpha}$ denote all the scalar fields and gaugino
fields, respectively.
The parameters in the mSUGRA model,  a universal scalar mass $m_0$,
a universal gaugino mass $m_{1/2}$, and the universal trilinear coupling
$A$ terms are introduced at the GUT scale:
\begin{equation}
m_i^2 = m_0^2, ~~~ m_{\alpha} = m_{1/2}, ~~~
A^{U,D,L} = A_0 Y^{U,D,L},
\end{equation}
where $Y^{U,D,L}$ are the diagonalized $3 \times 3$ Yukawa matrices.
In this model, there are four free parameters, $m_0$, $m_{1/2}$, $A_0$,
and $\tan \beta \equiv \langle H_2 \rangle / \langle H_1 \rangle$, in
addition to the sign of $\mu$.  The parameters $m_{1/2}$, $\mu$ and
$A$ can be complex, and four phases appear: $\theta_A$ (from $A_0$),
$\theta_1$ (from the gaugino mass $m_1$), $\theta_3$ (from the gaugino
mass $m_3$), and $\theta_{\mu}$ (from the $\mu$ term).

The mSUGRA model can not explain the large negative value of
$\sin(2 \phi_1)_{\phi K_s}$, because in this model, the only source
of flavor violation is in the CKM matrix, which can not provide a
sufficient amount of flavor violation needed for the $b \to s$
transition in the processes $B \to \phi K$ \cite{Arnowitt:2003ev}.
The minimal extension of the mSUGRA has been studied to solve the
large negative $\sin(2\phi_1)_{\phi K_s}$ in the context
of QCDF \cite{Arnowitt:2003ev}, or both large negative
$\sin(2\phi_1)_{\phi K_s}$ and large BR of $B \to \eta' K$ in the
context of NF \cite{khalil}.

We consider only non-zero (2,3) elements in $A$ terms as a simplest
extension of the mSUGRA model.  This new piece enhances the left-right
mixing of the second and third generation.
The $A$ terms with only non-zero (2,3) elements can be expressed as
\begin{equation}
A^{U,D} = A_0 Y^{U,D} + \Delta A^{U,D},
\end{equation}
where $\Delta A^{U,D}$ are $3 \times 3$ complex matrices and
$\Delta A^{U,D}_{ij} = \left| \Delta A^{U,D}_{ij} \right|
e^{i\phi^{U,D}_{ij}}$ with $\left| \Delta A^{U,D}_{ij} \right| = 0$
unless $(i,j)= (2,3)~{\rm or}~(3,2)$.
It is obvious that the mSUGRA model is recovered if $\Delta A^{U,D}=0$.

It has been noticed \cite{Arnowitt:2003ev} that the SUSY contribution
mainly affect the Wilson coefficients $C_{8g(7\gamma)}$ and
$\tilde C_{8g(7\gamma)}$ and these coefficients do not change the weak
annihilation effects arising from the SM calculation.
In our analysis, we consider all the known experimental constraints
on the parameter space of the model, as in Ref. \cite{Arnowitt:2003ev}.
Those constraints come from the radiative $B$ decay process
$B \to X_s \gamma$ ($2.2\times 10^{-4}<Br(B \to X_s \gamma)<4.5\times
10^{-4}$\cite{Alam,bsg}), neutron and electron electric dipole moments
($d_{n} < 6.3 \times 10^{-26} e~cm$, $d_{e} < 0.21 \times 10^{-26} e~cm$
\cite{pdg}), relic density measurements, $K^0 - \bar K^0$ mixing
($\Delta M_K = (3.490 \pm 0.006) \times 10^{-12}$ MeV \cite{pdg}),
LEP bounds on masses of SUSY particles and the lightest Higgs
($m_h \geq 114$ GeV).
From the experimental constraints, we find that $\theta_1 \approx 22^0$,
$\theta_3 \approx 30^0$, and $\theta_{\mu} \approx -11^0$.  For the
phase $\theta_A$, we set $\theta_A = \pi$.

%%%%%%%%%%%%%%%%%%%%%%%%%%%%%%%%%%%%%%%%%%%%%%%%%%%%%%%%%%%%%%%%%
%%%%%%%%%%%%%%%%%%%%%%%%%%%%%%%%%%%%%%%%%%%%%%%%%%%%%%%%%%%%%%%%%
\begin{table}
\caption{$\sin(2\phi_1)_{\phi K_s}$ (left column) and ${\cal B}(B^{\pm}
\to \phi K^{\pm}) \times 10^6$ (right column) at $\tan\beta=10$
for various values of the parameters $|A_0|$, $m_{1/2}$ and
$\left| \Delta A^D_{23(32)} \right|$.  The unit for these parameters
are in GeV. }
\begin{tabular}{c|c|c|c|c|c|c|c|c|c}
\hline \hline
$|A_0|$ & \multicolumn{2}{|c|}{$800$} & \multicolumn{2}{|c|}{$600$} &
\multicolumn{2}{|c|}{$400$} & \multicolumn{2}{|c|}{$0$}
& $\left| \Delta A^D_{23(32)} \right|$
\\ \hline
$m_{1/2}=300$ & $\ba{c} -0.55 \ea$ & $\ba{c} 9.9 \ea$ & $\ba{c} -0.57
\ea$ & $\ba{c} 9.2 \ea$ & $\ba{c} -0.55
\ea$ & $\ba{c} 9.1 \ea$ & $\ba{c} -0.54 \ea$ & $\ba{c} 8.1 \ea$
 & $66 - 74$
\\ \hline
$m_{1/2}=400$ & $\ba{c} -0.56 \ea$ & $\ba{c} 9.9 \ea$ & $\ba{c} -0.53
\ea$ & $\ba{c} 9.6 \ea$ & $\ba{c} -0.56
\ea$ & $\ba{c} 9.2 \ea$ & $\ba{c} -0.58 \ea$ & $\ba{c} 8.5 \ea$
 & $150 - 168$
\\ \hline
$m_{1/2}=500$ & $\ba{c} -0.37 \ea$ & $\ba{c} 9.9 \ea$ & $\ba{c} -0.39
\ea$ & $\ba{c} 9.9 \ea$ & $\ba{c} -0.42 \ea$
& $\ba{c} 10.0 \ea$ & $\ba{c} -0.43 \ea$ & $\ba{c} 8.1 \ea$
 & $244 - 256$
\\ \hline
$m_{1/2}=600$ & $\ba{c} -0.32 \ea$ & $\ba{c} 7.6 \ea$ & $\ba{c} -0.30
\ea$ & $\ba{c} 7.5 \ea$ & $\ba{c} -0.30 \ea$
& $\ba{c} 7.5 \ea$ & $\ba{c} -0.05 \ea$ & $\ba{c} 7.1 \ea$
 & $270 - 304$
\\ \hline \hline
\end{tabular}
\end{table}
%%%%%%%%%%%%%%%%%%%%%%%%%%%%%%%%%%%%%%%%%%%%%%%%%%%%%%%%%%%%%%%%%
%%%%%%%%%%%%%%%%%%%%%%%%%%%%%%%%%%%%%%%%%%%%%%%%%%%%%%%%%%%%%%%%%

%%%%%%%%%%%%%%%%%%%%%%%%%%%%%%%%%%%%%%%%%%%%%%%%%%%%%%%%%%%%%%%%%
%%%%%%%%%%%%%%%%%%%%%%%%%%%%%%%%%%%%%%%%%%%%%%%%%%%%%%%%%%%%%%%%%
\begin{table}
\caption{$\mathcal{A}_{b\to s + \gamma}$ (left column) and
$\mathcal{A}_{\phi K^{\pm}}$ (right column) at $\tan\beta=10$.}
\begin{tabular}{c|c|c|c|c|c|c|c|c}
\hline \hline
$|A_0|$ & \multicolumn{2}{|c|}{$800$} & \multicolumn{2}{|c|}{$600$} &
\multicolumn{2}{|c|}{$400$} & \multicolumn{2}{|c}{$0$}
\\ \hline
$m_{1/2}=300$ & $\ba{c} 0.011 \ea$ & $\ba{c} 0.21 \ea$ & $\ba{c}
0.017 \ea$ & $\ba{c} 0.22 \ea$ & $\ba{c}
0.02 \ea$ & $\ba{c} 0.22 \ea$ & $\ba{c} 0.024 \ea$ & $\ba{c} 0.23 \ea$\\
\hline  $m_{1/2}=400$ & $\ba{c} 0.017 \ea$ & $\ba{c} 0.21\ea$ & $\ba{c}
0.028 \ea$ & $\ba{c} 0.21 \ea$ & $\ba{c}
0.027 \ea$ & $\ba{c} 0.22 \ea$ & $\ba{c} 0.029\ea$ & $\ba{c} 0.23 \ea$\\
 \hline  $m_{1/2}=500$ & $\ba{c} 0.033 \ea$ & $\ba{c} 0.18 \ea$ & $\ba{c}
0.034 \ea$ & $\ba{c} 0.18 \ea$ & $\ba{c}
0.033 \ea$ & $\ba{c} 0.19 \ea$ & $\ba{c} 0.03 \ea$ & $\ba{c} 0.21 \ea$\\
 \hline  $m_{1/2}=600$ & $\ba{c} 0.023 \ea$ & $\ba{c} 0.2 \ea$ & $\ba{c}
0.023 \ea$ & $\ba{c} 0.2 \ea$ & $\ba{c}
0.023 \ea$ & $\ba{c} 0.2 \ea$ & $\ba{c} 0.02 \ea$ & $\ba{c} 0.16 \ea$\\
\hline \hline
\end{tabular}
\end{table}
%%%%%%%%%%%%%%%%%%%%%%%%%%%%%%%%%%%%%%%%%%%%%%%%%%%%%%%%%%%%%%%%%
%%%%%%%%%%%%%%%%%%%%%%%%%%%%%%%%%%%%%%%%%%%%%%%%%%%%%%%%%%%%%%%%%

%%%%%%%%%%%%%%%%%%%%%%%%%%%%%%%%%%%%%%%%%%%%%%%%%%%%%%%%%%%%%%%%%
%%%%%%%%%%%%%%%%%%%%%%%%%%%%%%%%%%%%%%%%%%%%%%%%%%%%%%%%%%%%%%%%%
\begin{table}[h]
\caption{${\cal B}(B^{\pm} \to \eta' K^{\pm}) \times 10^6$ (left column)
and ${\cal B}(B^{\pm}\rightarrow \eta K^{\pm}) \times 10^6$ (right column)
at $\tan\beta=10$.
\label{small4}}
\begin{tabular}{c|c|c|c|c|c|c|c|c}
\hline \hline
$|A_0|$ & \multicolumn{2}{|c|}{$800$} & \multicolumn{2}{|c|}{$600$} &
\multicolumn{2}{|c|}{$400$} & \multicolumn{2}{|c}{$0$}
\\ \hline
$m_{1/2}=300$ & $\ba{c} 79.6 \ea$ & $\ba{c} 3.67 \ea$ & $\ba{c} 81.0
\ea$ & $\ba{c} 3.76 \ea$ & $\ba{c} 79.6
\ea$ & $\ba{c} 3.66 \ea$ & $\ba{c} 79.0 \ea$ & $\ba{c} 3.62 \ea$\\
 \hline  $m_{1/2}=400$ & $\ba{c} 78.2 \ea$ & $\ba{c} 4.4 \ea$ & $\ba{c} 83.0
\ea$ & $\ba{c} 3.85 \ea$ & $\ba{c} 79.0
\ea$ & $\ba{c} 3.69 \ea$ & $\ba{c} 81.0 \ea$ & $\ba{c} 3.69 \ea$\\
 \hline  $m_{1/2}=500$ & $\ba{c} 84.8 \ea$ & $\ba{c} 3.90 \ea$ & $\ba{c}83.7
\ea$ & $\ba{c} 3.85 \ea$ & $\ba{c} 81.0 \ea$
& $\ba{c} 3.71 \ea$ & $\ba{c} 77.0 \ea$ & $\ba{c} 3.50 \ea$\\
 \hline  $m_{1/2}=600$ & $\ba{c} 73.0 \ea$ & $\ba{c} 3.26 \ea$ & $\ba{c} 71.0
\ea$ & $\ba{c} 3.18 \ea$ & $\ba{c} 70.0 \ea$
& $\ba{c} 3.10 \ea$ & $\ba{c} 70.0 \ea$ & $\ba{c} 3.00 \ea$\\
\hline \hline
\end{tabular}
\end{table}
%%%%%%%%%%%%%%%%%%%%%%%%%%%%%%%%%%%%%%%%%%%%%%%%%%%%%%%%%%%%%%%%%
%%%%%%%%%%%%%%%%%%%%%%%%%%%%%%%%%%%%%%%%%%%%%%%%%%%%%%%%%%%%%%%%%

%%%%%%%%%%%%%%%%%%%%%%%%%%%%%%%%%%%%%%%%%%%%%%%%%%%%%%%%%%%%%%%%%
%%%%%%%%%%%%%%%%%%%%%%%%%%%%%%%%%%%%%%%%%%%%%%%%%%%%%%%%%%%%%%%%%
\begin{table}[h]
\caption{$\mathcal{A}_{\eta' K^+}$ (left column) and $\mathcal{A}_{\eta K^+}$
(right column) at $\tan\beta=10$.
\label{small3}}
\begin{tabular}{c|c|c|c|c|c|c|c|c}
\hline \hline
$|A_0|$ & \multicolumn{2}{|c|}{$800$} & \multicolumn{2}{|c|}{$600$} &
\multicolumn{2}{|c|}{$400$} & \multicolumn{2}{|c}{$0$}
\\ \hline
$m_{1/2}=300$ & $\ba{c} 0.0 \ea$ & $\ba{c} -0.24 \ea$ & $\ba{c}
0.0 \ea$ & $\ba{c} -0.23 \ea$ & $\ba{c}
0.0 \ea$ & $\ba{c} -0.23 \ea$ & $\ba{c} -0.001 \ea$ & $\ba{c} -0.22\ea$\\
 \hline  $m_{1/2}=400$ & $\ba{c} -0.026 \ea$ & $\ba{c} -0.31 \ea$ & $\ba{c}
-0.003 \ea$ & $\ba{c} -0.22\ea$ & $\ba{c}
-0.004 \ea$ & $\ba{c} -0.24 \ea$ & $\ba{c} -0.003 \ea$ & $\ba{c} -0.23 \ea$\\
 \hline  $m_{1/2}=500$ & $\ba{c} -0.005 \ea$ & $\ba{c} -0.21 \ea$ & $\ba{c}
-0.005 \ea$ & $\ba{c} -0.21 \ea$ & $\ba{c}
-0.008\ea$ & $\ba{c} -0.22 \ea$ & $\ba{c} -0.002 \ea$ & $\ba{c} -0.22 \ea$\\
 \hline  $m_{1/2}=600$ & $\ba{c} -0.0004 \ea$ & $\ba{c} -0.22 \ea$ & $\ba{c}
-0.0006 \ea$ & $\ba{c} -0.22 \ea$ & $\ba{c}
-0.001 \ea$ & $\ba{c} -0.22 \ea$ & $\ba{c} -0.009 \ea$ & $\ba{c} -0.195 \ea$\\
\hline \hline
\end{tabular}
\end{table}
%%%%%%%%%%%%%%%%%%%%%%%%%%%%%%%%%%%%%%%%%%%%%%%%%%%%%%%%%%%%%%%%%
%%%%%%%%%%%%%%%%%%%%%%%%%%%%%%%%%%%%%%%%%%%%%%%%%%%%%%%%%%%%%%%%%

Now we consider the case with non-zero $\Delta A^D_{23}$ and
$\Delta A^D_{32}$ for $\tan \beta =10$ in our calculation (Similar
results can be obtained for other values of $\tan\beta$).  All the
other elements in $\Delta A^{U,D}$ are set to be zero. We also
implement $\left| \Delta A^D_{23} \right| \sim\left| \Delta
A^D_{32} \right|$ and $\phi^D_{23} \neq \phi^D_{32}$. In this
case, in order to find solutions, we do not need to use the
additional strong phase: i.e., $\delta' =0$. The value of
$m_{1/2}$ varies from 300 GeV to 600 GeV, and the value of $|A_0|$
varies from 0 to 800 GeV.  Even though the value of $m_0$ is not
explicitly shown, it is chosen for different $m_{1/2}$ and $A_0$
such that the relic density constraint is satisfied. (We satisfy
the relic density constraint using the stau-neutralino
co-annihilation channel \cite{cnst}.) The value of $m_0$ increases
as $m_{1/2}$ increases. The value of $\left| \Delta A^D_{23(32)}
\right|$ increases as $m_{1/2}$ does.  The phases $\phi^D_{23}$
and $\phi^D_{32}$ are approximately $-40^0$ to $-15^0$ and $165^0$
to $180^0$, respectively.

Table IV shows the BRs for $B^{\pm} \to \phi K^{\pm}$ (right column) and
$\sin(2\phi_1)_{\phi K_s}$ (left column), calculated for various values of
the parameters $m_{1/2}$ and $|A_0|$.
The value of $\sin(2\phi_1)_{\phi K_s}$ can be large and negative. If we
use the additional strong phase $\delta' \neq 0$, the magnitude of
$\sin(2\phi_1)_{\phi K_s}$ can be even larger.
For example, if $\delta' = -30^0$, then $\sin(2\phi_1)_{\phi K_s} = -0.65$
for $m_{1/2} =400$ GeV and $A_0 = -800$ GeV.
In Table V, we show the CP rate asymmetries $\mathcal{A}_{b \to s + \gamma}$
(left) and $\mathcal{A}_{\phi K^{\pm}}$ (right).  Since $A_{23}$ contributes
to $b \to s\gamma$, the CP asymmetry gets generated, but the asymmetry is
small.
So far we have assumed that $\Delta A^U_{23,32}=0$. But if we use $\Delta
A^U_{23,32}\neq 0$ and $\Delta A^D_{23,32}=0$, the value of
$\sin(2\phi)_{\phi K_s}$ is mostly positive.

The BRs for $B^{\pm} \to \eta' K^{\pm}$  and $B^{\pm} \to \eta K^{\pm}$
are shown in Table VI and the CP asymmetries are shown in Table VII.
They are consistent with the experimental data. The value of
$\sin(2\phi_1)_{\eta' K_s}$ is $0.6 - 0.7$ for all these scenarios.
The BRs are: $\mathcal{B}(B^0 \to \eta' K^{0}) \sim (63-73) \times 10^{-6}$
and $\mathcal{B}(B^0 \to \eta K^{0}) \sim (2.25-3.37) \times 10^{-6}$.
The CP asymmetries are:
$\mathcal{A}_{\eta' K^{0}} \sim \mathcal{A}_{\eta' K^+}$ and
$\mathcal{A}_{\eta K^{0}} \sim \mathcal{A}_{\eta K^+}$.
We see that the CP asymmetries are also in good agreement with the
experimental values shown in Table I.
As a final comment, we note that in the case of the large $X_{A,H}$ effects
with $\chi^2_{\min}=7.5$,
it is possible to raise the BR for $B^+ \to \phi K^+$ to $(8 - 9) \times
10^{-6}$.  However, in that case, $\mathcal{A}_{\rm CP}(B^+ \to \phi K^+)
\sim -0.2$ and $\sin(2\phi_1)_{\phi K_s} \geq -0.34$.

In conclusion, we have analyzed the CP asymmetries of $B\to \phi K$ and
$B\to \eta^{(\prime)} K$ modes in the QCDF framework. The phenomenological
parameters $X_H$ and $X_A$ arising from end-point divergences in the hard
spectator scattering and weak annihilation contributions are determined by
the global analysis for twelve $B \to PP$ and $VP$ decay modes, such as
$B \to \pi \pi$, $\pi K$, $\rho \pi$, $\rho K$, etc, but excluding the modes
whose (dominant) internal quark-level process is $b \to s \bar s s$.
We found that it is possible to explain large negative
$\sin(2\phi_1)_{\phi K_s}$ simultaneously with the large BRs for
$B^{\pm (0)} \to \eta' K^{\pm (0)}$ in the context of supersymmetry.
We use R-parity conserving SUGRA models and constrain the parameter space
with the observed dark matter relic density along with other experimental
constraints.  We also show that the R-parity violating SUSY models can
provide solutions.
Our results are consistent with other available data on CP asymmetry for
different $B \to \phi K$, $B \to \eta' K$, and $B \to \eta K$ modes.

%%%%%%%%%%%%%%%%%%%%%%%%%%%%%%%%%%%%%%%%%%%%%%%%%%%%%%
%%%%%%%%%%%%%%%%%%%%%%%%%%%%%%%%%%%%%%%%%%%%%%%%%%%%%%
\vspace{1cm}
\noindent
The work of B.D was supported by Natural Sciences and Engineering
Research Council of Canada. The work of C.S.K. was supported
in part by Grant No. R02-2003-000-10050-0 from BRP of the KOSEF and in part by
CHEP-SRC Program.
The work of S.O. and G.Z. was supported by the Japan Society for the
Promotion of Science (JSPS).

%%%%%%%%%%%%%%%%%%%%%%%%%%%%%%%%%%%%%%%%%%%%%%%%%%%%%%
%%%%%%%%%%%%%%%%%%%%%%%%%%%%%%%%%%%%%%%%%%%%%%%%%%%%%%

\end{document}